\documentclass[twocolumn,showpacs,preprintnumbers,amsmath,amssymb,floatfix]{revtex4}
\usepackage[dvips]{graphicx,epsfig,color}% Include figure files
\usepackage{dcolumn}% Align table columns on decimal point
\usepackage{bm}% bold math

\def\mdmatm{\Delta m^2_{32}}

\def\mdmsol{\Delta m^2_{21}}

\def\numunue{$\nu_\mu \rightarrow \nu_e$}

\def\meV{e\mbox{V}}

\begin{document}

\title{\large \bf Very Long Baseline Neutrino Oscillation Experiment\\ for Precise 
Measurements of Mixing Parameters\\ and  CP Violating Effects\\}

\author{ M.V.~Diwan, D.~Beavis, Mu-Chun~Chen, J.~Gallardo, R.L.~Hahn, S.~Kahn, H.~Kirk,  
W.~Marciano, W.~Morse, 
Z.~Parsa, 
N.~Samios, Y.~Semertzidis, B.~Viren, W.~Weng, P.~Yamin, M.~Yeh}
\affiliation{ Physics Department, Brookhaven National Laboratory, Upton, NY 11973}  

\author{W.~Frati, K.~Lande, A.K.~Mann, R.~Van~Berg, P.~Wildenhain}
\affiliation{Department of Physics and Astronomy, University of Pennsylvania, Philadelphia, 
PA 19104}

\author{J.R.~Klein}
\affiliation{Department of Physics,
University of Texas, Austin Texas 78712}

\author{I.~Mocioiu}
\affiliation{ Physics Department, University of Arizona, Tucson, AZ 85721}

\author{ R.~Shrock}
\affiliation{ C. N. Yang Institute for Theoretical Physics,
 State University of New York,
 Stony Brook, NY 11794}
 
\author{K.T.~McDonald}
\affiliation{Joseph Henry
Laboratories, Princeton University, Princeton, NJ 08544}

\baselineskip=11.6pt

\begin{abstract}
We analyze the prospects of a feasible, 
Brookhaven National Laboratory  based, very long baseline (BVLB) neutrino
oscillation experiment consisting of a conventional horn produced low
energy wide band beam and a detector of 500 kT fiducial mass with
modest requirements on event recognition and resolution.  Such an
experiment is intended primarily to determine CP violating effects 
in the neutrino sector for 3-generation mixing.
 We analyze the sensitivity of such an
experiment.  We conclude that this experiment 
will allow determination of the CP phase $\delta_{CP}$ and 
the currently unknown mixing parameter $\theta_{13}$, 
if  $\sin ^2 2 \theta_{13} \geq 0.01$,
a value  $\sim 15$ times lower than the present experimental upper
limit.  
In addition to  $\theta_{13}$ and $\delta_{CP}$, 
the experiment has great potential for precise
measurements of most other  
parameters in the neutrino mixing matrix including
$\Delta m^2_{32}$, $\sin^2 2\theta_{23}$, 
$\Delta m^2_{21}\times \sin 2 \theta_{12}$, and the
mass ordering of neutrinos through the observation of the matter
effect in the $\nu_\mu \to \nu_e$ appearance channel.

\end{abstract}

\baselineskip=14pt

\pacs{12.15.Ff,13.15.+g,14.60.Lm,14.60.Pq,11.30.Fs,11.30.Er}

\maketitle 

\section*{Introduction}

\noindent 
Measurements of solar and atmospheric neutrinos have provided strong
evidence for non-zero neutrino masses and mixing
\cite{ref1,sk,ref2}.  Atmospheric results have been further
strengthened by the K2K collaboration's accelerator based results
\cite{k2k}.  The Solar neutrino results have been confirmed by the
KamLAND collaboration in a reactor based experiment that has shown
that the large mixing angle (LMA) solution is most likely the correct
one \cite{kamland}.  Interpretation of the experimental results is
based on oscillations of one neutrino flavor state, $\nu_e, \nu_{\mu}$
or $\nu_{\tau}$, into the others, and described quantum mechanically
in terms of  neutrino mass eigenstates, $\nu_1,
\nu_2$ and $\nu_3$. The mass squared differences  involved in
the transitions are measured to be approximately $\Delta m^2_{21}
\equiv  m {(\nu_2)}^2 - m {(\nu_1)}^2  = (5 -20) \times 10^{-5} {\rm
eV}^2$ for the solar neutrinos and $\Delta m^2_{32} \equiv
m{(\nu_3)}^2 - m{(\nu_2)}^2 = \pm(1.6 - 4.0) \times 10^{-3} {\rm eV}^2$
for the atmospheric neutrinos, with large mixing strengths, $\sin^2 2
\theta_{12}\approx 0.86$ and $\sin^2 2 \theta_{23}\approx 1.0$ in both
cases. 

These existing data on neutrino oscillations\cite{ref1,sk,ref2} and
the prospect of searching for CP violation make clear that the next
generation of oscillation experiments must be significantly more
ambitious than before. In particular, the source of neutrinos needs to
be accelerator based so that both the neutrino flavor content and the
energy spectrum of the initial state can be selected. Several
approaches  have been explored in the literature. These involve
either a narrow band ``off-axis'' beam produced with a conventional magnetic
focusing system \cite{e889,offaxis,barger} or a neutrino factory
based on a muon storage ring \cite{geer}. In this paper, we show that
the currently favored parameters open the possibility for an
accelerator based very long baseline (BVLB) experiment that can explore both
solar and atmospheric oscillation parameters in a single experiment,
complete the measurement of the mixing parameters, and search for new
physics.

\section*{The Experimental Strategy}

There are four  measurements of primary interest that can be addressed
with the experimental arrangement described in this paper. Using the
parameter convention described in \cite{ref1}, they are:

\begin{enumerate}
\item[] (i) Definitive observation of oscillatory behavior 
(multiple oscillations) in 
the $\nu_{\mu}$ disappearance mode and precise determination of the 
oscillation parameters $\Delta m^2_{32}$ and $\sin^2 2 \theta_{23}$, 
with statistical errors $\sim \pm 1\%$. 

\item[] (ii) Detection of the oscillation $\nu_{\mu} \rightarrow \nu_e$ 
in the appearance mode and measurement of the parameter $\sin^2 2
\theta_{13}$. This will involve matter enhancement and also allow
definitive measurement of the sign of $\Delta m^2_{32}$, i.e., which
neutrino $\nu_3$ or $\nu_2$ is heavier.

\item[] (iii) Measurement of the CP violation parameter, $\delta_{CP}$,
 in the lepton sector.

\item[] (iv) Measurement of $\Delta m^2_{21}\times \sin 2 \theta_{12}$ 
in the appearance channel of $\nu_\mu \to \nu_e$.

\end{enumerate}

\begin{figure}
  \begin{center}
    \includegraphics*[width=0.48\textwidth]{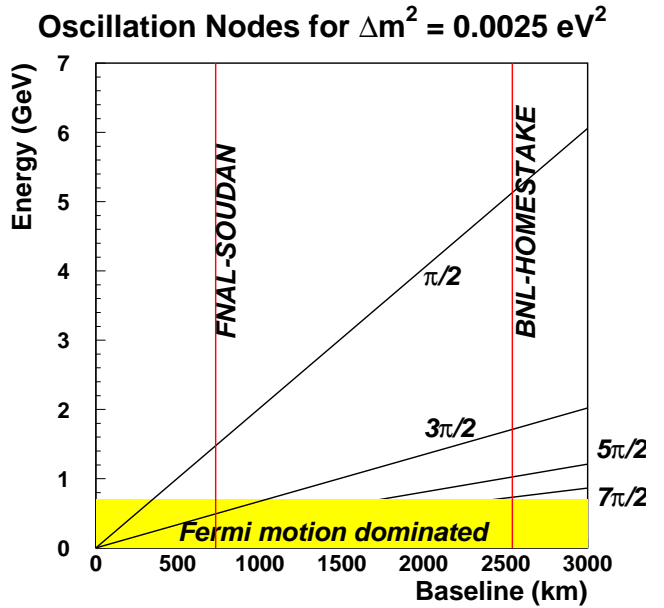} 
    \caption[Oscillation nodes {\it vs.} distance.]
{(color) Nodes of neutrino oscillations for disappearance (not
affected by matter effects) as a function of oscillation length and
energy for $\Delta m^2_{32} = 0.0025 \ {\rm eV}^2$. The distances from
FNAL to Soudan  and
from BNL to Homestake are shown by the vertical lines.}
    \label{nodes}  
  \end{center}
\end{figure}

We describe how these measurements could be carried out
with good precision in a single accelerator based experiment.
% making
%use of the already measured oscillation parameters and under
%reasonable assumptions for as yet unmeasured ones.
For precise and definitive measurement of oscillations we must observe
multiple oscillation nodes in the energy spectrum of reconstructed muon and
electron neutrino charged current events. The multiple node signature
is also necessary in order to distinguish between oscillations and
other more exotic explanations such as neutrino decay
\cite{ndecy} or extra dimensions \cite{barbieri}  for the muon 
neutrino deficit in atmospheric neutrinos. Since the cross-section,
Fermi motion, and nuclear effects limit the statistics and the energy
resolution (for reconstructed neutrino energy) of low energy charged
current events, we use neutrinos with energies greater than few
hundred MeV and clean events with a single visible muon or electron
that are dominated by quasielastic scattering, $\nu_l + n \to l + p$,
where $l$ denotes a lepton.  Figure \ref{nodes} shows that the
distance needed to observe at least 3 nodes in the reconstructed
neutrino energy spectrum is greater than 2000 km for $\Delta m^2_{32}
= 0.0025 {\rm ~eV^2}$, the currently favored value from
Super-Kamiokande atmospheric data \cite{sk}.  A baseline of longer
than 2000 km coupled with a wide band beam with high flux from 0.5 to
6 GeV will provide a nodal pattern in the $\nu_\mu \to \nu_\mu$
disappearance channel and good sensitivity over a broad range of
$\Delta m^2_{32}$.

\begin{figure}
  \begin{center}
    \includegraphics*[width=0.48\textwidth]{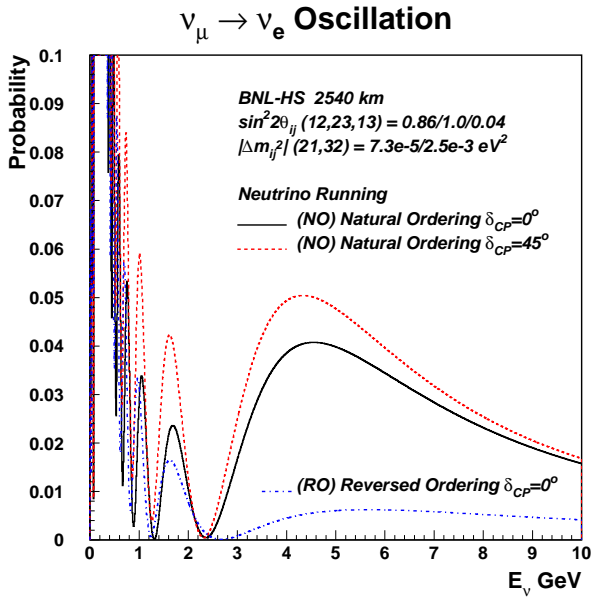}
    \caption[P($\nu_\mu \to \nu_e$).]{(color) Probability of \numunue{}
        oscillations at 2540 km.
The calculation includes the effects of matter. 
The dotted ($\delta_{CP} = 45^o$) 
and solid  ($\delta_{CP} = 0^o$) curves 
are for {\it NO} and the
lower dot-dashed ($\delta_{CP} = 0^o$) 
curves is for {\it RO}.
The parameters used for  the figure are 
$\sin^2 2 \theta_{12}=0.86$, $\sin^2 2 \theta_{23}=1.0$, and 
$\sin^2 2 \theta_{13}=0.04$ and 
$\Delta m^2_{21}=7.3\times 10 ^{-5}~eV^2$, 
$\Delta m^2_{32}=0.0025~eV^2$.
    }
    \label{app1}
  \end{center}
\end{figure}

The low energy wide band beam and the very long baseline also 
have a number
of important advantages for the appearance channel of $\nu_\mu \to
\nu_e$. These advantages can be summarized using Figure \ref{app1}
which shows the probability of $\nu_\mu \to \nu_e$ oscillation as a
function of neutrino energy for the distance of 2540 km, the distance
between BNL and the Homestake mine in South Dakota. The oscillation
parameters that we have assumed are indicated in the figure and the
caption. We define the natural mass ordering ({\it NO}) of neutrinos
to be $m_3 > m_2 > m_1$, and the other two possibilities ($m_2> m_1 >
m_3$ and $m_1 > m_2 > m_3$) are collectively called reversed ordering
({\it RO}). Both {\it RO} possibilities have very similar physical
consequences, but $m_1 > m_2 > m_3$ is strongly disfavored by the
explanation of the Solar neutrino deficit using matter enhanced
neutrino oscillation (the LMA solution).  Since neutrinos from an
accelerator beam must pass through the Earth to arrive at a detector
located 2540 km away, the probability in Figure
\ref{app1} includes the effects of matter, which enhance (suppress) the
probability above 3.0 GeV for {\it NO} ({\it RO}) \cite{ref12}.
 Therefore
the appearance probability above 3.0 GeV is sensitive to both the mass
ordering and the parameter $\sin^2 2 \theta_{13}$. The probability in
the region 1.0 to 3.0 GeV is less sensitive to matter but much more
sensitive to the CP phase $\delta_{CP}$ as shown in Figure
\ref{app1}\cite{ref3}.  The increase in the probability below 1.5 GeV
is due to the presence of terms involving the solar oscillation
parameters, $\Delta m^2_{21}$ and $\sin^2 \theta_{12}$.  Therefore,
the spectrum of electron neutrino events measured with a wide band
beam over 2540~km with sufficiently low background has the potential
to determine $\sin^2 2\theta_{13}$, $\delta_{CP}$, and 
the mass ordering of neutrinos as well as
$\Delta m^2_{21}$ because these parameters affect different regions of
the energy spectrum.  In the following, we  examine how well the
parameters can be determined and the implications for the detector
performance and background.

\section*{Proposed Experimental Configuration}

The high energy proton accelerator, to be used for making the neutrino
beam, must be intense ($\sim$ 1 MW in power) to provide sufficient
neutrino-induced event rates in a massive detector very distant from
it.  Such a long baseline experimental arrangement \cite{ref4} may be
realized with a neutrino beam from the upgraded 28 GeV proton beam
from the Alternating Gradient Synchrotron at the Brookhaven National
Laboratory (BNL) and a water Cherenkov detector with 0.5 megaton of
fiducial mass at either the Homestake mine in South Dakota or the
Waste Isolation Pilot Plant (WIPP) in New Mexico, at distances $\sim$2540 km
and $\sim$2900 km from BNL, respectively.  A version of the detector
is described at length in \cite{3m} and is a candidate for location in
the Homestake Mine, where it would occupy five independent cavities,
each about 100 kton in fiducial volume. Another version of the
detector in a single volume is described in
\cite{uno}.
For much of the discussion below we have used the distance of 2540 km
to Homestake for our calculations. Although the statistics obtained 
at 2900 km to WIPP will be somewhat smaller, many of the physics 
effects will be larger, and therefore the resolution and the physics 
reach of the BVLB experiment are approximately independent of 
the choice between Homestake or WIPP.

The accelerator upgrade as well as the issues regarding the production
target and horn system are described in \cite{agsup}. We briefly
describe it here for completeness. Currently, the BNL-AGS can
accelerate $\sim 7\times 10^{13}$ protons upto 28 GeV approximately
every 2 sec.  This corresponds to average beam power of about 0.16 MW.
This average power could be upgraded by increasing the repetition rate
of the AGS synchrotron to 2.5 Hz while keeping the number of protons
per pulse approximately the same.  Currently a 200 MeV room
temperature LINAC in combination with a small synchrotron, called the
Booster, injects protons into the AGS at 1.2 GeV. The process of
collecting sufficient number of protons from the Booster into the AGS
takes about 0.6 sec. Therefore, for 2.5 Hz operation the Booster must
be replaced by a new injector. A new super-conducting LINAC to replace
the Booster could serve the role of a new injector; the remaining
modifications to the AGS are well understood and they involve power
supplies, the RF system, and other rate dependent systems to make the
accelerator ramp up and down at 2.5 Hz. We expect the final upgraded
accelerator configuration to yield $8.9\times 10^{13}$ protons every
400 ms at 28 GeV. Our studies 
 show that this intensity can be further 
upgraded to 2 MW by incremental improvements  to the LINAC and the 
AGS repetition rate.

In addition to the 1 MW accelerator upgrade, the pion production
target, the horn focusing system, and the decay tunnel, which must be
aimed at an angle of 11.26 degrees into the earth (towards Homestake)
need to be built. Our current studies indicate that a
carbon-carbon 
composite target inserted into a conventional aluminum
horn cooled with forced Helium and water will function in the 1 MW
beam for a sufficient length of time.  Our current plan is to build a
hill instead of an underground tunnel to accommodate the 200 m long
pion decay tunnel.  We consider both the rapid cycling accelerator
upgrade which maintains a relatively low intensity per pulse and the
1 MW capable target technically less risky than other alternatives.
For comparison, at JPARC \cite{jparc} the first (preliminary) stage of 0.75 
MW is to be achieved
at $\sim$ 1/3 Hz with $3.3\times 10^{14}$ protons per pulse at 50 GeV.
At Fermilab a 2 MW upgrade to the 120 GeV main injector calls for
either an 8 GeV super-conducting LINAC or a new rapid cycling proton
synchrotron as injectors for the main injector\cite{pdriver}.  
As explained in the
following, the combination of the very long baseline and a 500 kT
detector allows the BVLB experiment
 to use a lower average power at 1 MW compared to
the proposed JPARC or NuMI(Off axis) \cite{offaxis} 
 projects and achieve  wider physics
goals. This reduces the technical difficulties (liquid mercury or
moving targets, radiation damage, shielding of personnel, etc.) 
associated with using beam powers in excess of 1 MW.

\begin{figure}
  \begin{center}
    \includegraphics*[width=0.48\textwidth]{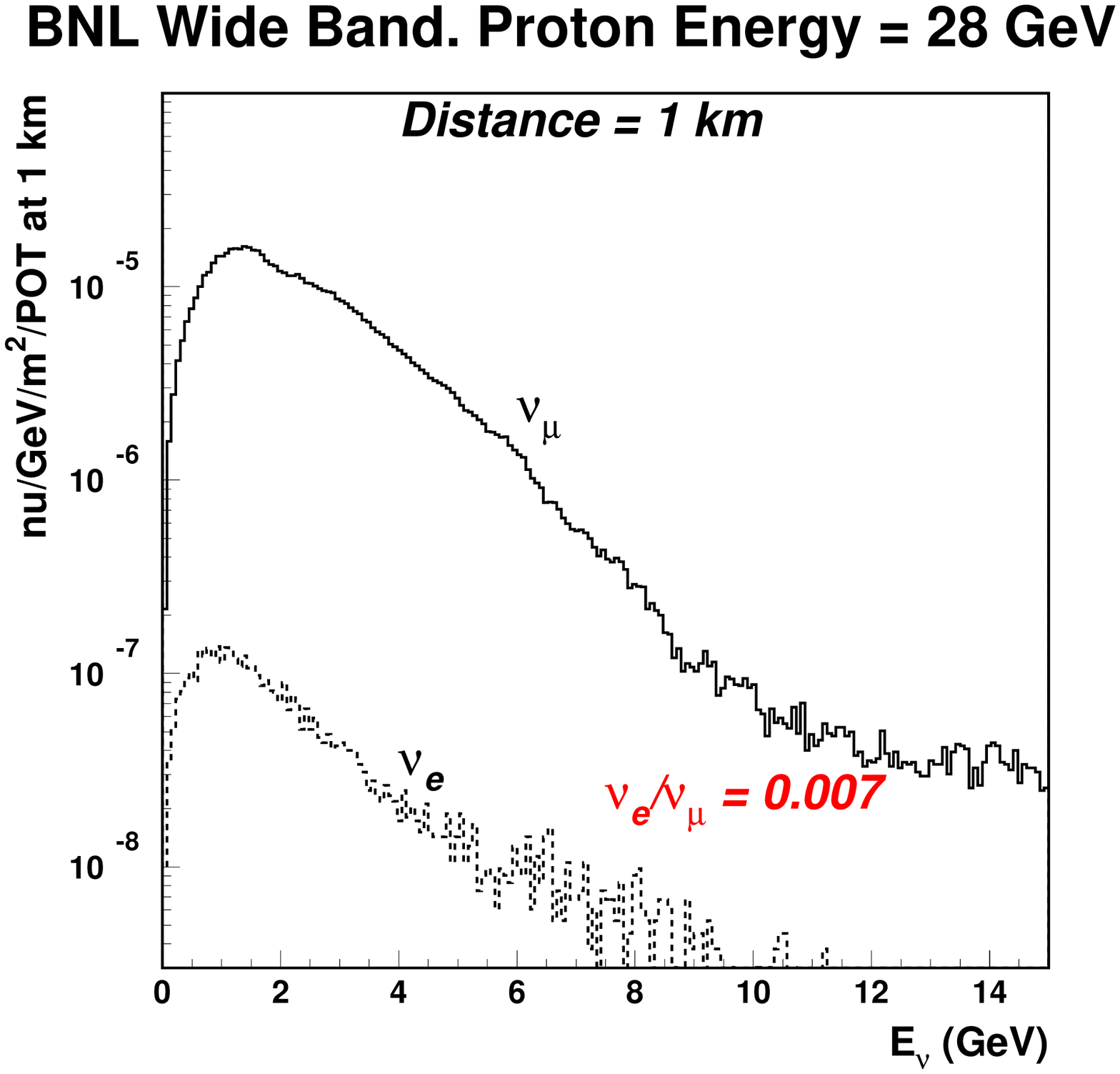} 
    \caption[Wide band neutrino spectrum from 28 GeV protons.]
{(color) The simulated 
wide band neutrino flux for 28 GeV protons on a graphite 
target  used 
for the calculations in this paper. (POT$=$protons on target).
}
    \label{nuspec}  
  \end{center}
\end{figure}

We have performed detailed simulations of a wide band horn focused
neutrino beam using 1 MW of 28 GeV protons on a graphite target. The
neutrino spectrum obtained in these simulations and used for the
results in this paper is shown in Figure \ref{nuspec} normalized for
POT (protons on target). We calculate that with such a beam (total
intensity of $\sim 4.7\times 10^{-5}~\nu/m^2/POT$ at distance of 1 km
from the target) we will obtain about 60,000 charged current and
20,000 neutral current events in a total exposure lasting $5\times
10^7$ seconds in a 0.5 megaton water Cherenkov detector located at the
Homestake mine 2540 km away from BNL.
%We expect to get
%this running time in 4 months of running each year for 5 years.  
%We use the spectra obtain by making various cuts to 
% appropriate for a water
%Cherenkov detector to select single muon or electron events 
% for the results reported below.
For the results reported below, we use the particle spectra obtained
by making cuts to select single muon or electron quasielastic events.
Events with multiple particles (about 4 times higher in statistics) 
could be used to further enhance the
statistical significance of the effects.

A conventional horn focused beam can be run in either the neutrino or
the anti-neutrino mode. In this paper we show that most of the physics
program in the case of 3-generation mixing can be carried out by
taking data in the neutrino mode alone. It is well known that
experiments that must take data with anti-neutrinos, which have lower
cross section, must run for long times to accumulate sufficient
statistics \cite{offaxis}. This problem is largely alleviated in our
method. We will discuss the anti-neutrino beam in a separate paper;
here we wish to focus on the physics reach of the neutrino beam alone
\cite{antinu}.

\section*{\boldmath$\nu_{\mu}$ Disappearance}

\begin{figure}
  \begin{center}
  \includegraphics*[width=0.48\textwidth]{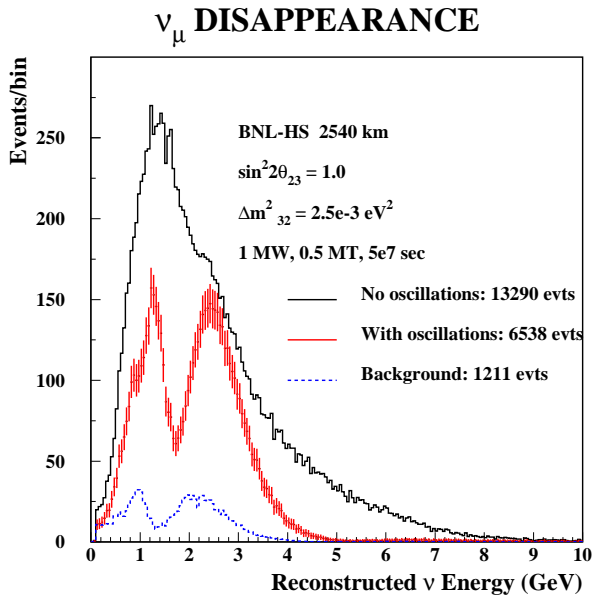}
  \caption[Expected $\nu_\mu$ disappearance spectra, $\Delta m^2_{32}
  = 0.0025$] {(color) Spectrum of expected single muon events in a 0.5 MT
  water Cherenkov detector at 2540 km from the source.  We have
  assumed 1 MW of beam power and $5 \times 10^7$ sec of
  data-taking. The top histogram is without oscillations; the middle
  histogram with error bars is with oscillations. Both histograms
  include the dominant single pion charged current background. The
  bottom histogram shows this background contribution to the
  oscillating spectrum.  This plot is for $\Delta m^2_{32} = 0.0025 \
  {\rm ~eV}^2$. The error bars correspond to the statistical error
  expected in the bin. At low energies the Fermi movement, which is
  included in the simulation, will dominate the resolution. } 
  \label{wcnodesa} \end{center}
\end{figure}

\begin{figure}
  \begin{center}
  \includegraphics*[width=0.48\textwidth]{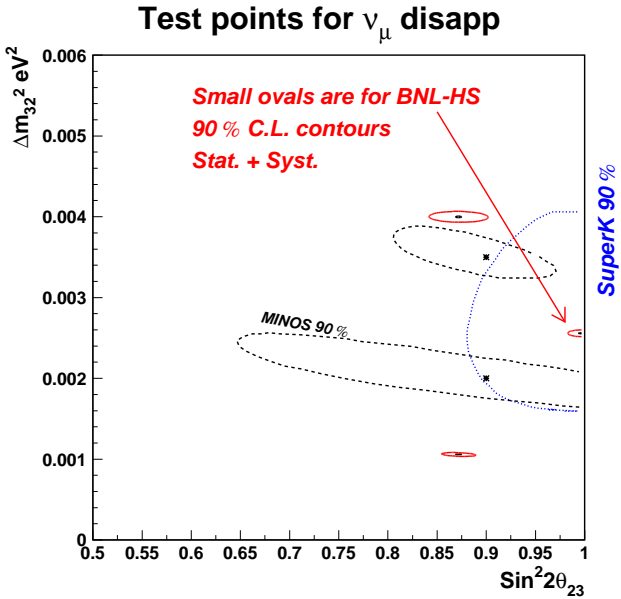}
  \caption[Statistical and systematic uncertainty for $\Delta
  m^2_{32}$ and $\sin^22\theta_{23}$, includes allowed regions from
  Super Kamiokande and expected resolution from MINOS.]  { (color) 
Resolution
  including statistical and systematic effects at 90\% confidence
  level on $\Delta m_{32}^2$ and $\sin^2 2\theta_{23}$ for the 2540
  baseline experiment; assuming 1 MW, 0.5 MT, and $5\times 10^7$ sec
  of exposure.  We have included a 5\% bin-to-bin systematic
  uncertainty in the energy calibration as well as a 5\% systematic
  uncertainty in the normalization.  We have not included a systematic
  uncertainty on the global energy scale; this should be added in
  quadrature to the expected resolution on $\Delta m^2_{32}$. The
  expected resolution from the MINOS experiment at Fermilab (dashed)
  and the allowed region from Super Kamiokande 
(dotted) 
is also indicated. } 
  \label{cntr} \end{center}
\end{figure}

An advantage of the very long baseline is that the multiple node
pattern is detectable over the entire allowed range of $\Delta m^2_{32}$.
 The sensitivity to $\Delta m^2_{32} \sim 1.24
\frac{E_{\nu}{\rm [GeV]}}{L\rm [km] } ~eV^2 $
 extends to the low value of about $5 \times 10^{-4} {\rm eV}^2$, 
significantly below the allowed range from Super Kamiokande. 
 At energies lower than $\sim 1$ GeV, the $\nu_{\mu}$ energy resolution
will be dominated by Fermi motion and nuclear effects, as illustrated
in Figure \ref{nodes}. The contribution to the resolution from water
Cherenkov track reconstruction depends in first approximation on the
photo-multiplier tube coverage. With coverage greater than 10\%, a
reconstruction energy resolution of better than $\sim 10\%$ should be
achieved \cite{e889}. 
 The simulated spectrum of the expected $\nu_{\mu}$
disappearance signal including backgrounds 
and resolution is shown in Figure
\ref{wcnodesa} for $\Delta m^2_{32} = 0.0025 {\rm ~eV^2}$ as a
function of reconstructed neutrino energy. The background, which will
be primarily charged current events, will also oscillate and 
broaden the dips in the nodal pattern.  From this spectrum we 
estimate that the determination of $\Delta
m^2_{32}$ will have a statistical uncertainty of approximately $\pm
0.7$\% at $\Delta m^2_{32} = 0.0025\, {\rm ~eV}^2$ 
and $\sin^2 2 \theta_{23}= 1$. 
The experiment can determine $\sin^2 2\theta_{23} > 0.99$ at 90\%
confidence level.  Within the parameter region of interest there will
be little correlation in the determination of $\Delta m^2_{32}$ and
$\sin^2 2\theta_{23}$. The precision of the experiment is compared in
Figure \ref{cntr} with the precision expected from   MINOS
\cite{minos} and the result from 
 Super-Kamiokande \cite{sk}.
Since the statistics and the size of the expected signal (distortion
of the spectrum) are both large in the disappearance measurement, we
expect that the final error on the parameters will be dominated by the
systematic error.  A great advantage of the very long baseline and
multiple oscillation pattern in the spectrum is that the effect of
systematic errors on  flux normalization, background subtraction,
and spectrum distortion due to nuclear effects or detector
calibration is small. The error on the overall detector energy scale
is expected to be the dominant systematic error
\cite{ref4}. The large event rate in this experiment will allow 
us to measure $\Delta m^2_{32}$ precisely in a short period of time;
this measurement will be  important to predict the shape of the 
appearance signal which we  now discuss.

\section*{\boldmath$ \nu_{\mu} \rightarrow \nu_e$ Appearance}

The importance of matter effects on long-baseline neutrino oscillation
experiments has been recognized for many years \cite{ref12,arafune}.
For our study, we have included the effect of matter with a full
numerical calculation taking into account a realistic matter profile in the
earth, following  \cite{ref11}. 
For our present
discussion, it is useful to exhibit an approximate analytic formula for the
oscillation of $\nu_\mu \to \nu_e$ for 3-generation mixing obtained with the
simplifying assumption of constant matter density \cite{freund, cervera}. 
Assuming a constant matter density, 
the oscillation of $\nu_{\mu} \rightarrow \nu_e$ in the Earth 
for 3-generation mixing is described
approximately by  Equation \ref{qe1}.
In this equation $\alpha=\Delta m^2_{21}/\Delta m^2_{31}$, $\Delta = \Delta
m^2_{31} L/4E$, $\hat{A}=2 V E/\Delta m^2_{31}$,
$V=\sqrt{2} G_F n_e$. $n_e$ is the density of electrons in the Earth. 
Recall that $\Delta m^2_{31} = \Delta m^2_{32}+\Delta m^2_{21}$. 
Also notice that $\hat{A}\Delta = L G_{F} n_e/\sqrt{2}$ is sensitive only 
to the sign of $\Delta m^2_{31}$. 

\begin{widetext}
\begin{eqnarray}
P(\nu_{\mu} \rightarrow \nu_e) &\approx&
\sin^2 \theta_{23} {\sin^2 2 \theta_{13}\over (\hat{A}-1)^2}\sin^2((\hat{A}-1)\Delta)  \nonumber\\ &&
+\alpha{\sin\delta_{CP}\cos\theta_{13}\sin 2 \theta_{12} \sin 2
\theta_{13}\sin 2 \theta_{23}\over
\hat{A}(1-\hat{A})} \sin(\Delta)\sin(\hat{A}\Delta)\sin((1-\hat{A})\Delta) \nonumber\\ &&
+\alpha{\cos\delta_{CP}\cos\theta_{13}\sin 2 \theta_{12} \sin 2
\theta_{13}\sin 2 \theta_{23}\over
\hat{A}(1-\hat{A})} \sin(\Delta)\cos(\hat{A}\Delta)\sin((1-\hat{A})\Delta) \nonumber\\ &&
+\alpha^2 {\cos^2\theta_{23}\sin^2 2 \theta_{12}\over
\hat{A}^2}\sin^2(\hat{A}\Delta) \nonumber \\
\label{qe1}
\end{eqnarray}
%\end{widetext}

For anti-neutrinos, the second term in Equation \ref{qe1}  
has the opposite sign. It  is
proportional to the following CP violating quantity.

%\begin{widetext}
\begin{equation}
J_{CP} \equiv \sin\theta_{12} \sin\theta_{23} \sin\theta_{13} \cos\theta_{12} 
\cos\theta_{23} \cos^2\theta_{13} \sin \delta_{CP}
\label{eq2}
\end{equation}
\end{widetext} 

Equation \ref{qe1}
is an  expansion in powers of $\alpha$. 
The  approximation becomes inaccurate 
 for $\Delta m^2_{32} L/4E > \pi/2$ as well as $\alpha \sim 1$. 
For the actual
results shown in Figure \ref{app1} we have used the exact numerical
calculation, accurate to all orders. 
Nevertheless, the approximate formula is
useful for understanding important features of the appearance probability:
1) the first 3 terms in the equation control the enhancement (for {\it NO})
or suppression (for {\it RO}) of the oscillation 
probability above 3 GeV; 
2) the second and third terms control the sensitivity to
CP in the intermediate 1 to 3 GeV range; and 3) the last term controls
the sensitivity to $\Delta m^2_{21}$ at low energies. 

%If the values of
%$\Delta m^2_{31}$ and $\Delta m^2_{21}$ are such that both contribute to
%$\nu_\mu \to \nu_e$ with approximately equal strength then the above
%approximation cannot be used ($\alpha << 1$ is not valid) and large
%interference effects in the very long baseline experiment will be
%seen.  

While the $\nu_{\mu}$ disappearance result will be affected
mainly by systematic errors, the $\nu_{\mu} \rightarrow \nu_{e}$
appearance result will be affected  by the backgrounds. The $
\nu_e$ signal will consist of clean, single electron events (single
showering rings in a water Cherenkov detector) that result mostly 
from the
quasielastic reaction $\nu_e + n \rightarrow e^- + p$. The main
backgrounds will be from the electron neutrino contamination in the
beam and reactions that have a $\pi^0$ in the final state. The $\pi^0$
background will depend on how well the detector can distinguish events
with single electron induced and two photon induced 
electromagnetic (e.m.) showers.
Backgrounds due to $\nu_\mu \to \nu_\tau$ conversion,
$\nu_\tau$ charged current reactions in the detector, and subsequent  
$\tau^- \to e^- \bar\nu_e \nu_\tau$ decays are small because of the 
low energy of the beam and the consequent  low cross section for $\tau$
production \cite{ref4}.

\begin{figure}
  \begin{center}
  \includegraphics*[width=0.48\textwidth]{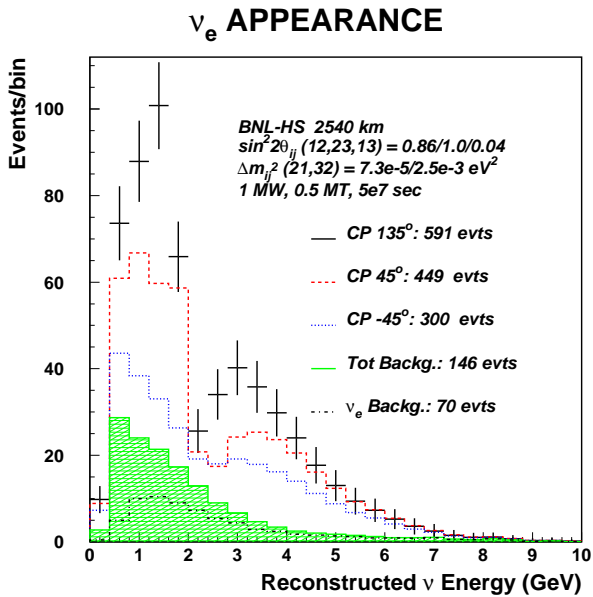} \\
  \caption[Expected spectrum of reconstructed electron neutrino
  energy.]  {(color) The expected electron neutrino spectrum for 3 
  values of the CP parameter $\delta_{CP}$ including background
  contamination. The points with error bars are for $\delta_{CP} = 135^0$; the
  error bars indicate the expected statistical error on each bin. The histogram
  directly below the error bars is for $\delta_{CP} = 45^0$ and the
  third histogram is for $\delta_{CP} = -45^0$. The hatched histogram
  shows the total background. The $\nu_e$ beam background is also
  shown. The plot is for $\Delta m^2_{32} = 0.0025
\ {\rm eV}^2$. We have assumed $\sin^2 2 \theta_{13} = 0.04$ and
$\Delta m^2_{21} = 7.3 \times 10^{-5} \ {\rm eV}^2$. The values of
$\sin^2 2 \theta_{12}$ and $\sin^2 2 \theta_{23}$ are set to 0.86, 1.0,
respectively. Running conditions as in Figure \ref{wcnodesa}.  } 
\label{app2} \end{center}
\end{figure}

Because of the rapid fall in the BVLB neutrino spectrum beyond 4
GeV (Figure \ref{nuspec}), the largest contribution to the $\pi^0$
background will come from neutral current events with a single $\pi^0$
in the final state. It is well known that resonant single pion
production in neutrino reactions has a rapidly falling cross section
as a function of momentum transfer, $q^2$, \cite{adler} up to the
kinematically allowed value. This characteristic alone suppresses this
background by more than 2 orders of magnitude for $\pi^0$ (or shower)
energies above 2 GeV
\cite{ref4}. 
Therefore a modest $\pi^0$ background suppression (by a factor of
$\sim 15$ below 2 GeV and $\sim$2 above 2 GeV) is sufficient to reduce
the $\pi^0$ background to manageable level over the entire spectrum.
The electron neutrino contamination in the beam from decays of muons
and kaons, is well understood to be 0.7\% of the muon neutrino flux
\cite{e734beam}, with 
a similar spectrum (Figure \ref{nuspec}).
  The experimentally observed
electron neutrino energy 
spectrum will therefore have three components: the
rapidly falling shape of the $\pi^0$ background, the spectral
 shape of
the $\nu_e$ beam contamination
slightly modified by oscillation, and the oscillatory shape of the
appearance signal. The shape of the appearance spectrum will be
 well known because of the precise knowledge of $\Delta m^2_{32}$
from the disappearance measurement as well as the improved
 knowledge of $\Delta
m^2_{21} $ from KamLAND \cite{kamland}. These distinguishing spectra
will  allow experimental detection of $\nu_\mu \to \nu_e$ with
good confidence. Figure \ref{app2} shows a simulation of the
expected spectrum of reconstructed electron neutrino energy 
after $5\times 10^7$ sec 
 of running. The parameters assumed are listed in the figure as
well as the caption.

Figure \ref{app2} further illustrates the previously described three
regions of the appearance spectrum: the high energy region ($> 3$ GeV)
with a matter enhanced (for {\it NO}) appearance has the main
contribution to the background from $\nu_e$ contamination in the beam;
the intermediate region ($1 \to 3$ GeV) with high sensitivity to the
CP phase, but little dependance on mass ordering, has approximately
equal contribution from both background sources
($\nu_e$ and $\pi^0$);  and the low energy
region ($<1.5$ GeV) , where the effects of the CP phase and $\Delta
m^2_{21}$ dominate, has the main background from the $\pi^0$
events.  Matter enhancement of the oscillations has been postulated
for a long time without experimental confirmation \cite{ref12}.
Detection of such an effect by observing a matter enhanced peak around
3 GeV will be  important. However, in the case of {\it RO} mass
ordering this enhanced peak will be missing, but the effect (depending
on $\delta_{CP}$) on the rest of the electron neutrino spectrum will
be small.

\begin{figure}
  \begin{center}
  \includegraphics*[width=0.48\textwidth]{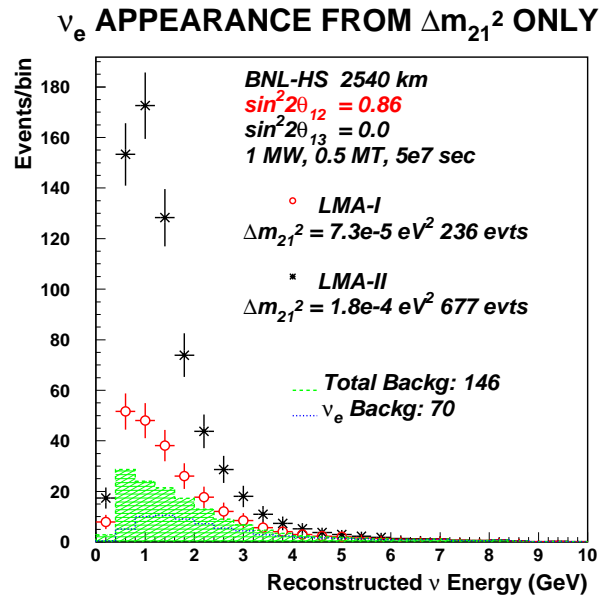} \\
  \caption[Expected spectrum of reconstructed electron neutrino energy
  for $\theta_{13}=0$.]  {(color) The expected spectrum of electron neutrino
  events for $\sin^2 2 \theta_{13} = 0$. Important parameters in this
  figure are $\Delta m^2_{21} = 7.3 \times 10^{-5} \ {\rm eV}^2$ (LMA-I)
or  $\Delta m^2_{21} = 1.8 \times 10^{-4}$ (LMA-II)
   and
  $\sin^2 2 \theta_{12} = 0.86$. 
 All other parameters and the running
  condition as in Figure \ref{wcnodesa}.  } \label{app3} \end{center}
\end{figure}

The very long baseline combined with the low
energy spectrum make it posible to observe $\nu_\mu \to \nu_e$
conversion even if $\sin^2 2 \theta_{13}= 0$ because of the contribution
from $\Delta m^2_{21}$ if the solar neutrino large mixing angle
solution (LMA) holds. In Figure \ref{app3} we show the expected
appearance spectrum for $\Delta m^2_{21} = 7.3\times 10^{-5}eV^2$
(LMA-I) with 90 excess events
and $\Delta m^2_{21} = 1.8\times 10^{-4}eV^2$ (LMA-II)
with 530 excess events above background. 
The LMA-I and LMA-II solutions 
 with their respective confidence levels are 
discussed in \cite{fogli}. 
This measurement will be
sensitive to the magnitude and knowledge of the background because
there will be no oscillating signature to distinguish the signal.  We
estimate that the statistical and systematic errors on this
measurement will allow us to determine 
the combination 
$\Delta m^2_{21}\times sin 2 \theta_{12}$ to a
precision of about 10\% at the LMA best fit point. This is competitive
with the expected measurement from KamLAND. However, it will be done
in the explicit ($\nu_\mu \to \nu_e$) 
 appearance mode, which is qualitatively different from the
measurements made so far in the SNO or the KamLAND experiments.
Therefore, such a measurement will have important implications on 
the search for new physics such as sterile neutrinos.

\begin{figure}
  \begin{center}
  \includegraphics*[width=0.48\textwidth]{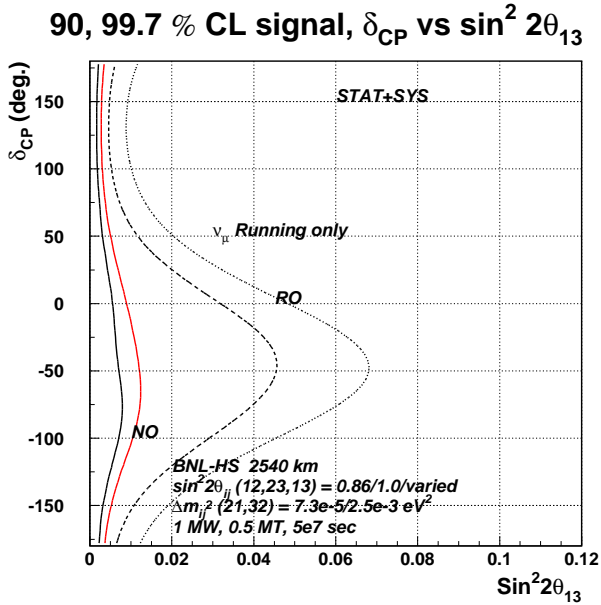} \\
  \caption[ 90, and 99.73\% C.L. contours on $\sin^2 2 \theta_{13}$
  versus $\delta_{CP}$.  ] {(color)
 90 and 99.73\% C.L. contours of the BVLB 
  experiment in the variables $\sin^2 2 \theta_{13}$ versus
  $\delta_{CP}$ for the natural ({\it NO}) and reversed  ordering of
  parameters ({\it RO}). The solid lines are for {\it NO}, the left
  line for 90\% and right line for 99.73\% C.L. The dashed and dotted
  lines are for 90 and 99.73\% C.L. for {\it RO}. 
  On the right hand side of the 99.73\% C.L. lines the experiment
  will observe an electron appearance signal with greater than 3 sigma
  significance. } \label{app4} \end{center}
\end{figure}

In Figure \ref{app4} we show the 90 and 99.73\% (3 sigma)
C.L. sensitivity of the BVLB experiment in the variables $\sin^2 2
\theta_{13}$ versus $\delta_{CP}$. The actual limit obtained in the
case of a lack of signal will depend on various ambiguities. Here we
show the 99.73\% C.L. lines for {\it NO} and {\it RO}, on the right
hand of which the experiment will observe an electron appearance
signal due to the presence of terms involving $\theta_{13}$ with
greater than 3 sigma significance and thus determine the corresponding
mass ordering. Currently from atmospheric data there is a 
 sign uncertainty on $\theta_{23} = \pm\pi/4$; this 
introduces an additional ambiguity onto Figure \ref{app4} of
$\delta_{CP} \to \delta_{CP} +\pi$.  For this plot we have assumed
that the other parameters are well known either through other
experiment or by the disappearance measurement: $\Delta m^2_{21}
= 7.3 \times 10^{-5} {\rm eV}^2$, $\Delta m^2_{32} = 0.0025 {\rm
~eV^2}$.  The values of $\sin^2 2\theta_{12}$ and $\sin^2 2
\theta_{23}$ are set to 0.86, 1.0, respectively. 
  The sensitivity to $\sin^2 2 \theta_{13}$ is approximately constant
  in the $\Delta m^2_{32}$ range allowed by Super Kamiokande
\cite{ref4}. The value of $\Delta m^2_{21}$ affects the
modulation of the $\sin^2 2 \theta_{13}$ sensitivity with respect to
$\delta_{CP}$.  If there is no excess electron appearance signal,
other than the expected signal due to $\Delta m^2_{21}$ alone, then it
would indicate a switch to anti-neutrino 
running to determine if the
{\it RO} hypothesis with parameters on the left hand side of the
dashed line in Figure \ref{app4} applies \cite{antinu}.

\section*{Sensitivity to the CP Violation Parameter}

To get a qualitative understanding for the  measurement 
of CP violating parameters we compare the size of
terms involving $\delta_{CP}$ with the first term in Equation
\ref{qe1}. This ratio is seen to be approximately proportional to  
$\alpha \sin \delta_{CP} (\Delta m^2_{21} L/4 E_\nu)$.  As shown in
Figure \ref{app1}, the fractional contribution from the CP violating
terms increases for lower energies at a given distance. 
The energy dependence of the CP effect and the matter effect 
tend to be opposite and therefore can be distinguished from each other 
from the energy distribution using neutrino data alone.  
On the other hand,  the statistics for a given size detector at a
given energy are poorer by one over the square of the distance, but
the term linear in $\sin \delta_{CP}$ grows linearly in distance
\cite{ref3}. The statistical sensitivity (approximately proportional
to the square root of the event rate) to the effects of CP violation,
therefore, is independent of distance because the loss of event rate
and the increase of the CP effect approximately cancel each other
in the statistical merit  if
backgrounds remain the same.  Therefore, the two important advantages
of the BVLB approach are that the CP effect can be
detected without running in the anti-neutrino mode and the sensitivity
to systematic errors on the background and the normalization is
considerably reduced because the fractional size of the CP effect is
large. For example, in Figure
\ref{app1}, the CP effect at $\delta_{CP} = \pi/4$ in the first
oscillation peak is $\sim 20\%$ while the effect in the second
oscillation peak is more than 50\%. Therefore, it is unnecessary to
know the background and the normalization to better than 10\% to
obtain a significant measurement of $\delta_{CP}$
at the second oscillation maximum.
There could be a  contribution to the systematic error  from the
theoretical calculation of the probability shown in Figure
\ref{app1}.  
Because of the very long baseline, this probability depends on the
Earth's density profile which is known to about 5\%.  Random density
fluctations on that order will lead to a relative systematic
uncertainty in the $\nu_e$ appearance probability of about 1\%
\cite{mattersys}, which is not significant for the BVLB method, 
but could be significant in
the case of an experiment that performs the CP measurement at the
first oscillation maximum with a shorter baseline.

\begin{figure}
  \begin{center}
    \includegraphics*[width=0.48\textwidth]{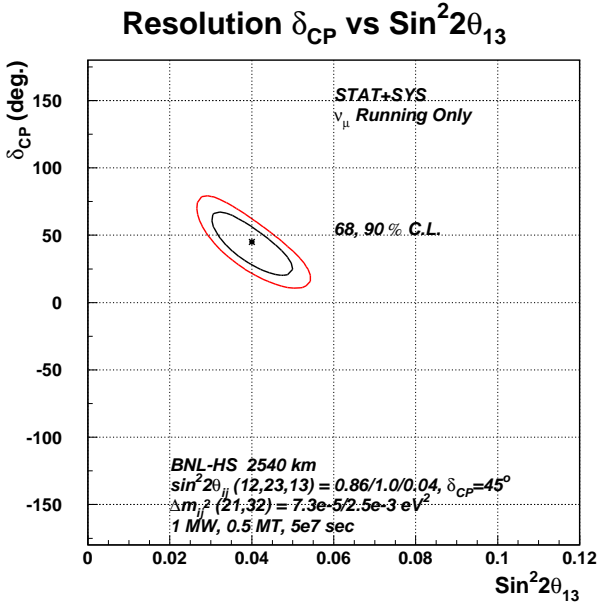}
    \caption[Expected  statistical + systematic uncertainties at test point of $\delta_{CP} = 45^\circ$ and $\sin^22\theta_{13} = 0.04$] 
{(color) 68\% and 90\% confidence level error contours in  $\sin^2  2 \theta_{13}$ 
versus $\delta_{CP}$ for statistical and systematic errors. 
The test point used here is  
$\sin^2  2 \theta_{13}=0.04$ and $\delta_{CP}=45^o$.
 $\mdmatm = 0.0025 ~\meV^2$, and   $\mdmsol = 7.3\times 10^{-5} ~\meV^2$. The values of 
$\sin^2 2 \theta_{12}$ and $\sin^2 2 \theta_{23}$ are set to 
0.86, 1.0, respectively.  }    
    \label{app5}
  \end{center}
\end{figure}

\begin{figure}
  \begin{center}
    \includegraphics*[width=0.48\textwidth]{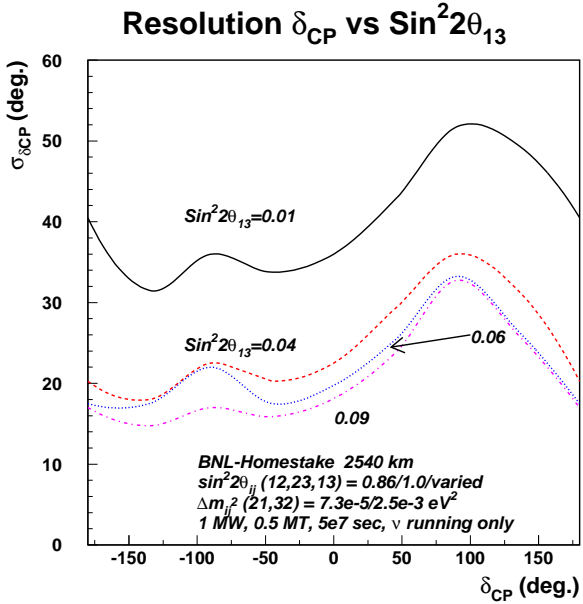} \\
    \caption[Resolution on $\delta_{CP}$ versus $\delta_{CP}$]
{(color) 1 sigma error on $\delta_{CP}$ in degrees for several values of 
$\sin^2 2 \theta_{13}$. All other parameters are held fixed assuming 
they will be known before this measurement.
    }
    \label{app6}
  \end{center}
\end{figure}

An experiment with a wide band
beam is  needed to 
 exploit the energy dependence of the CP effect 
to extract the
value of $\delta_{CP}$ and $\sin^2 2 \theta_{13} $ 
using only the neutrino data. 
The  energy dependence also reduces the correlation between 
the two parameters.  
%correlation between the two parameters is much reduced 
%because the effect of $\delta_{CP}$ has an energy dependence 
%opposite to that of $\sin^2 2 \theta_{13}$. 
Figure 
\ref{app5} shows the expected resolution on $\sin^2 
2 \theta_{13}$
versus $\delta_{CP}$ at a particular case, 
$\sin^2 2 \theta_{13}=0.04$ and 
$\delta_{CP}=\pi/4$ with all other parameters fixed  as indicated.
The 1$\sigma$ error on $\delta_{CP}$ from this measurement
as a function of $\delta_{CP}$ is shown in Figure \ref{app6}.  It is
clear that  the precision of the measurement will be limited 
if  $\sin^2 2
\theta_{13} < 0.01$ because of the background.
 However, for larger values the
resolution on $\delta_{CP}$ is approximately independent of $\sin^2 2
\theta_{13}$. 
This result can be understood by examining Equation \ref{qe1} in which 
the first term grows as $\sin^2 2
\theta_{13}$ and the CP violating term grows as $\sin 2 \theta_{13}$.
Therefore the  statistical sensitivity to the CP violating 
term, which is the ratio of the second term to 
the square-root of the first term, is indenpendent of $\theta_{13}$
as long as the background does not dominate \cite{ref3}.  
At $\delta_{CP}\approx 135^o$ the event rate reaches
maximum and does not change rapidly with respect to $\delta_{CP}$,
therefore the resolution on $\delta_{CP}$ becomes  poor.

Because of the multiple node spectrum the appearance
probability is sensitive to both the $\sin \delta_{CP}$ and $\cos
\delta_{CP}$ terms in Equation \ref{qe1}. This eliminates the
$\delta_{CP} \to (\pi-\delta_{CP})$ ambiguity 
normally present in a
narrow band single node experiment \cite{barger}. The sign uncertainty
$\theta_{23}=\pm \pi/4$ introduces a $\delta_{CP} \to (\delta_{CP} +
\pi)$ ambiguity. If $\Delta m^2_{21}$ is known poorly, there
will be a contribution to the 
 $\delta_{CP}$ resolution due to the 
correlation between $\delta_{CP}$ and $\Delta m^2_{21}$.  We
expect to know $\Delta m^2_{21}$ from KamLAND with precision of
$<10\%$ \cite{kamland}; this does not introduce a dominant source of
uncertainty on $\delta_{CP}$ in our experimental method. 
If a  four parameter, $\Delta m^2_{21}$, $\sin^2 2 \theta_{12}$,
$\sin^2 2 \theta_{13}$, and $\delta_{CP}$, 
 fit to the data is performed, the correlations between the parameters
 dilute the sensitivity to each parameter.
However,  the ambiguities and correlations  do not reduce the ability
of the experiment to establish that 3-generation neutrino mixing contains a
non-zero complex phase and hence a CP~violating term. This is seen if
we consider the resolution on the quantity $\Delta m^2_{21} \times
J_{CP}$ which is CP~violating and by definition does not exhibit any
of the above correlations. The measurement of this quantity will
simply depend on the statistics in the low  and medium energy
components  of the spectrum.

%An explicit
%demonstration of CP violation will require data taking with an
%anti-neutrino beam.  Such a demonstration could be carried out after
%the initial neutrino data taking period to confirm that there are no
%non-standard sources of CP violation in the neutrino sector beyond the
%mixing matrix.

%For small $\delta_{CP}$ the
%resolution on $\delta_{CP}$ will vary linearly with 
%respect to $\Delta m^2_{21}$ and therefore the 
%error on $\Delta m^2_{21}$ will contribute to the 
%final error on $\delta_{CP}$. 
%On the other hand, the resolution on the quantity 
%$\Delta m^2_{21} \times J_{CP}$ 
%and therefore will have a 
%weak (approximately square-root) dependence on $\Delta m^2_{21}$.

\section*{Comparison to Other Techniques}  
  
The alternatives to the BVLB plan  fall into two
categories: experiments with conventional horn focused beams over
baselines of several hundred km and with beams based on neutrinos from
muon decays in a high intensity muon storage ring. For a recent review
of muon storage ring based neutrino factory physics as well as comparisons
to conventional neutrinos beams see \cite{apollonio}.

Recently, for the conventional beam an ``off-axis'' beam is preferred
because the energy spectrum is narrower and can be tuned, within a
limited range from 1 to 2 GeV by choosing the off-axis angle
\cite{e889}.  A narrow energy spectrum could be better suited for
identifying $\nu_\mu \to \nu_e$ conversion because the 
neutral current backgrounds from interactions of 
 high energy neutrinos are lowered. The JPARC to Super Kamiokande project
(baseline of 295 km) has adapted this strategy and a new initiative at
Fermilab based on the NuMI beamline (baseline of 732 km) is also
considering it \cite{offaxis}. Both of these projects focus on the
observation of a non-zero value for $\theta_{13}$. Since it is not
possible to eliminate the $\sim$0.5\% $\nu_e$ contamination in a
conventional $\nu_\mu$ beam (even if it is ``off-axis'') the
sensitivity of all  projects based on a conventional beam, 
including the BVLB plan, 
is limited to $\sin^2 2 \theta_{13} \sim 0.005$ (ignoring the
effect of $\delta_{CP}$). 
Sensitivity below this value is background limited and 
requires very good understanding of
the systematic errors associated with the purity of the beam. 
The BVLB plan does, however, extend this sensitivity to values of 
$\Delta m^2 $ down to 0.001 $eV^2$.  
 The ``off-axis'' projects must be upgraded to
several Megawatts of beam power and detectors of several hundred kT
(the second phase of the JPARC project calls for 4 MW beam and a 1 megaton
detector) to perform the CP measurement. The narrow band nature of the
beam requires running in the anti-neutrino mode to observe CP
violation. The BVLB plan  uses a wide band beam over a much
longer distance to achieve the same sensitivity to $\sin^2 2
\theta_{13}$ over a wider range of $\Delta m^2_{32}$. It also has the
ability to see a multiple node structure in both the disappearance and
appearance channel; thus making the observation of $\nu_\mu \to \nu_e$
experimentally more robust against unanticipated backgrounds.  
In addition, the unique aspect of the wide band beam over the very long baseline
 is the ability to measure $\delta_{CP}$ with few
ambiguities using neutrino data alone.

The muon storage ring based ``neutrino factory'' solves the background
problem faced with conventional beams. A $\bar\nu_\mu$ and a $\nu_e$
is produced in each $\mu^+$ decay. An appearance search looks for the
conversion of $\nu_e$ into $\nu_\mu$ by looking for events with
negatively charged $\mu^-$ particles in the presence of $\bar\nu_\mu$
charged current events with positively charged $\mu^+$
particles in the final state.  
Such a physics signature should potentially have very low
background. Therefore, a neutrino factory based $\nu_e \to
\nu_\mu$ search is estimated to be sensitive to $\sin^2 2
\theta_{13}$ perhaps as small as $10^{-4} -- 10^{-5}$. 
Such a facility is very well tuned for measurement of $\sin^2 2
\theta_{13}$. Nevertheless, because of the relatively narrow energy
spectrum and higher energy of the neutrino factory, it lacks the
ability to observe multiple nodes in the observed spectrum.  For the
neutrino factory the effects of CP violation are observed by
comparing $\bar\nu_e \to \bar\nu_\mu$ and $\nu_e \to \nu_\mu$ in two
different runs of $\mu^-$ and $\mu^-$ decays in the presence of very
large matter effects.  The narrow spectrum also limits the ability of
the neutrino factory to resolve the ambiguities in the $\delta_{CP}$
determination regardless  of the value of $\sin^2 2 \theta_{13}$. This
has been extensively discussed in the literature \cite{barger,
cervera} and one of the proposed solutions is to construct several
detectors with different baselines to uniquely determine $\delta_{CP}$
versus $\sin^2 2 \theta_{13}$.  
Although the neutrino factory can extend
the reach for $\sin^2 2 \theta_{13}$ to very small values because the
backgrounds to the appearance search are small, the resolution on
$\delta_{CP}$ -- disregarding the ambiguities inherent in the 
neutrino factory approach -- is comparable to the BVLB resolution. 
We therefore conclude, that if $\sin^2 2 \theta_{13} \ge
0.01$ the BVLB experiment is either equal to or exceeds the physics reach
for neutrino mixing parameters of the neutrino factory.

Other suggestions in the literature include $\nu_e$ or $\bar\nu_e$
beams from either decays of radioactive nuclei accelerated to high
energies (beta-beam) \cite{beta} or $\bar\nu_e$'s from nuclear power
generation reactors such as the beam used for 
KamLAND \cite{kamland}. Both of these
concepts aim to create very pure, but  low energy beams.   For CP
sensitivity both $\nu_e$ and $\bar\nu_e$ must be produced in separate
runs in the beta-beam concept.  The low background should enable the
beta-beam concept to have good sensitivity to $\theta_{13}$, if 
sufficient flux can be obtained.  The
reactor beams are below $\nu_\mu$ charged current reaction threshold
and therefore must rely on disappearance to search for a non-zero
value of $\theta_{13}$. A disappearance based search for $\theta_{13}$ will
most likely be limited by systematic errors. 
Further comments on these techniques must await more detailed studies.

\section*{\bf Conclusions}

We have simulated and analyzed a feasible, very long baseline  neutrino
oscillation experiment consisting of a low energy, wide band neutrino
beam produced by 1 MW of 28 GeV protons incident on a carbon target
with magnetic horn focusing of pions and kaons, and a 500 kT detector
at a distance of $>$ 2500 km from the neutrino source. The BVLB neutrino
beam with a total intensity of about $4.7\times 10^{-5} \nu/m^2/POT$
at a distance of 1 km from the target could be provided by an upgrade
to the BNL-AGS \cite{agsup}.

 The single BVLB experiment could produce measurements of all parameters
 in the neutrino mixing matrix through observation of the
 disappearance channel, $\nu_{\mu} \rightarrow \nu_{\mu}$ and the
 appearance channel, $\nu_{\mu} \rightarrow \nu_e$.  
The experiment is also sensitive to the mass
 ordering of neutrinos using the observation of the matter effect in
 the appearance channel through the currently unknown parameter
 $\sin^2 2 \theta_{13}$. Nevertheless,  the experiment is intended
 primarily to measure the strength of CP~invariance violation in the
 neutrino sector and will provide a measurement of the CP~phase,
 $\delta_{CP}$ or alternatively the CP~violating quantity, $J_{CP}$, 
  if the one currently unknown neutrino oscillation
 mixing parameter $\sin ^2 2 \theta_{13} \geq 0.01$, a value about 15
 times lower than the present experimental upper limit. We
 point out that for a given resolution on $\delta_{CP}$ the number of
 neutrino events needed which determines the detector size and beam
 intensity, is approximately independent of the baseline as well as
 the value of $\sin ^2 2 \theta_{13}$ as long as the electron signal
 is not background dominated. Therefore,
the concept of very long baseline ($\ge
 2000~km$) is  attractive because it provides access to much
 richer physics phenomena as well as reduces the need to understand
 systematic errors on the flux normalization and background
 determination. A shorter baseline ($<1500 ~km$) will obviously limit
 the reach of the experiment for $\Delta m^2_{21}\times \sin 2
 \theta_{12}$ in the appearance mode as well as reduce the 
resolution on the CP parameter. On the other hand, a much longer
 baseline ($>4000 ~km$) will result in a matter effect that is large
 enough to dominate the spectrum and make the extraction of the
 CP~effect more difficult. The larger distance will also make it
 necessary to make the neutrino beam  directed
 at a higher angle into the Earth, a technical challenge that may not be  
necessary given the current values of neutrino parameters.
Lastly, we have also shown that most of this rich physics
 program including the search for CP~effects can be carried out by
 neutrino running alone. Once CP~effects are established with neutrino
 data, anti-neutrino data could be obtained for more precision on the
 parameters or search for new physics. The shape of the 
disappearance signal over multiple oscillations
from   neutrino running alone (as well as in combination with 
anti-neutrino running) can be used to contrain effects of new physics
from: sterile neutrino mixing, extra dimensions, exotic interactions
with matter, etc.

 It has not escaped our notice that the large detector size ($\ge$500
 kT), mandated by the above described neutrino program, naturally
 lends itself to the important physics of proton decay, supernova 
detection, etc. Although some
 experimental requirements such as shielding from cosmic rays,
 detector shape, or photo-sensor coverage, may differ, it is clear
 that they can be resolved so that this large detector can also be employed
 to address these frontier physics issues.

\section{Acknowledgements}

We thank all contributors to the BNL neutrino working group,
as well as 
 the BNL-AGS technical staff and engineers whom we consulted
during this work. We especially thank Thomas Roser, Nick Tsoupas,
Alessandro Ruggiero, Deepak Raparia, Nick Simos, and Hans Ludewig for
their contribution to the accelerator and beam design.  We also thank
 Thomas Kirk who encouraged us throughout this work.  This work was
supported by DOE grant DE-AC02-98CH10886 and NSF grant PHY02-18438.

\end{document}